\documentclass[12pt]{article}
\usepackage{a4,epsfig}
\begin{document}




\begin{titlepage}

\pagenumbering{arabic}
\vspace*{-1.5cm}
\begin{tabular*}{15.cm}{lc@{\extracolsep{\fill}}r}
&
\\
& &
14 June 2016
\\
&&\\ \hline
\end{tabular*}
\vspace*{2.cm}
\begin{center}
\Large 
{\bf \boldmath
 Experimental aspects  \\
of the tachyon hypothesis} \\
\vspace*{2.cm}
\normalsize { 
   
   {\bf V.F. Perepelitsa}\\
   {\footnotesize ITEP, Moscow            }\\ 
}
\end{center}
\vspace{\fill}
\begin{abstract}
\noindent
This note is aimed at an
introductory presentation of experimental characteristics of charged tachyons, 
deduced under the assumption that the tachyons possess standard electromagnetic
interactions. In particular, Cherenkov radiation by tachyons, their ionization 
loss in tracking devices and the bremsstrahlung loss in electromagnetic 
calorimeters are considered.

\end{abstract}
\vspace{\fill}

\vspace{\fill}
\end{titlepage}






\section{Introduction}
\setcounter{equation}{0}
\renewcommand{\theequation}{3.\arabic{equation}}
Currently an analysis of events containing anomalous Cherenkov rings 
is under progress using the data collected by the DELPHI experiment at LEP
during the LEP1 and LEP2 periods. The term ``anomalous rings" is related to
the rings of angular radius larger than $cos^{-1}(1/n)$
(where $n$ is the refractive index of the Cherenkov radiator).
A straightforward interpretation of such rings can be done 
within a framework of a hypothesis of faster-than-light particles, 
i.e. particles with spacelike momenta, called tachyons. 
This note is devoted to a description of some experimental
properties of these particles which are expected when considering charged 
tachyons possessing standard electromagnetic interactions.

The first theoretical arguments for the possibility of the existence of 
particles with spacelike momenta can be found in a famous paper by Wigner
in which the classification of unitary irreducible representations (UIR's)
of the Poincar\'{e} group was done for the first time~\cite{wigner1}.
In the 1960's Wigner returned to discuss the UIR's of the Poincar\'{e} 
group corresponding to particles with spacelike momenta \cite{wigner2}. He has
shown that quantum mechanical equations corresponding to these UIR's
describe particles with imaginary rest mass moving faster than light.
This almost coincided in time with the appearance of two pioneering works in
which the hypothesis of faster-than-light particles was formulated explicitly, 
accompanied by a kinematic description of them \cite{bds} 
(see Sect.~2) and by their quantum field theory \cite{fein}. 
The particles were called $tachyons$, from the Greek word 
$\tau \alpha \chi \iota \sigma$ meaning $swift$ \cite{fein}. 

These propositions immediately encountered strong objections 
related to the causality principle. It has been shown in several 
papers \cite{newton,roln,parment}, in agreement with an earlier remark by 
Einstein \cite{ein} (see also \cite{tolman,moller,bohm}), 
that by using tachyons as information carriers one can build a causal loop, 
making possible information transfer to the past time of an observer. 
This is deduced from the apparent ability of tachyons 
to move backward in time, which happens when they have 
a negative energy provided by a suitable Lorentz transformation, this property 
of tachyons being a consequence of the spacelikeness of their four-momenta.
A consensus was achieved that within the special relativity 
faster-than-light signals are incompatible with the principle of causality. 

Another important problem related to tachyons was their vacuum instability.
It is a well-known problem which usually appears when considering theoretical
models with a Hamiltonian containing a negative mass-squared term (for an
instructive description of the problem see e.g. \cite{nielsen}). Applied 
in a straightforward manner to consideration of faster-than-light particles
it results in a maximum, instead of a minimum, of the Hamiltonian for tachyonic
vacuum fields, and leads to the conclusion that the existence of tachyons 
as free particles is not possible.

Fortunately, both problems turned out to be mutually connected, and having 
hidden loopholes, they were resolved in the 1970's - 1980's, 
as described in detail in \cite{pvcaus,ttheor}.

In brief, the causality problem was resolved by combining the tachyon hypothesis
with the modern cosmology, which establishes the preferred reference frame:
so called comoving frame, in which the distribution of matter in the universe, 
as well the cosmic background (relic) radiation, are isotropic. This changes
the situation with the causality violation by tachyons drastically, since
the fast tachyons needed for a construction of a causal loop (they are called 
transcendent tachyons) are extremely sensitive to this frame. Therefore
this frame has to be involved when considering the propagation of tachyon
signals through space. These signals turn out to be ordered by the retarded 
causality in the preferred frame, and after the causal ordering being 
established in this frame, no causal loops appear in any other frame.    

Furthermore, in parallel with the causal ordering of the tachyon propagation 
one succeeds to get a stable tachyon vacuum which presents the minimum of the 
field Hamiltonian and appears, in the preferred frame, to be an ensemble
of zero-energy, but finite-momentum, on-mass-shell tachyons propagating
isotropically. The boundaries of this vacuum confine the acausal tachyons.

Several important properties of tachyons relevant for their experimental
characteristics were deduced in \cite{ttheor} from the general consideration 
of the infinite-dimensional UIR's of the O(2,1) subgroup of the Poincar\'{e} 
group, corresponding to faster-than-light particles. According to 
\cite{ttheor}, tachyons appear as extended, axially-symmetric, stringlike 
objects. Their spins are directed along their momenta, to be more properly
defined as helicities, always non-zero, since spinless (scalar) tachyons
cannot be realizations of these UIR's. Tachyons and antitachyons, by definition,
have positive and negative helicities, respectively, which may be either 
integer or half-odd-integer \cite{wigner2}. Tachyons can only be produced 
in pairs with antitachyons. The production of a tachyon of a high helicity 
state back-to-back with an antitachyon (generally speaking, with essentially
non-zero opening angle) is suppressed in any reaction by the angular momentum 
conservation, unless the antitachyon is produced in a direction parallel
to that of the tachyon, with the angular momenta of both particles compensating
(or almost compensating) each other. The overall angular momentum of such a 
pair can be low (even zero).

Our conclusion is that no fully convincing arguments can be raised against the
tachyon hypothesis. However we also conclude that in spite of a large progress 
made in the understanding of tachyon properties since the tachyon hypothesis 
first appeared, no complete and irrefutable theory describing them has so far
been formulated. Therefore our predictions for tachyon 
behaviour in an experimental set-up will be based on the fragmentary, very 
often semi-intuitive and semi-quantitative approaches.
 
This note is organized as follows. In Section~2 several kinematic formulae 
for tachyons are presented. Section~3 deals with the Cherenkov radiation of
charged tachyons. Characteristic dependence of the tachyon ionization loss 
on the tachyon velocity is sketched in Section~4. The behaviour of the charged
tachyons in electromagnetic calorimeters is described in Section~5, and
a comment on such a behaviour in transition radiation detectors is given in
Section~6. Section~7 contains a summary of the note. 

In formulae which follow below the velocity of light $c$
and the Planck constant $\hbar$ are taken to be equal to 1.
      
\section{Tachyon kinematics}
\setcounter{equation}{0}
\renewcommand{\theequation}{2.\arabic{equation}}
Faster-than-light particles were postulated in \cite{bds} possessing the 
following properties. They cannot traverse the light barrier and be
brought to rest in any reference frame. Therefore their rest mass is imaginary,
$m = i\mu$, mass squared is negative, $m^2 = -\mu^2$, which determines their
four-momentum, $P = (E,{\bf p})$ to be spacelike, $P^2 < 0$. Namely,
$E^2 - {\bf p}^2 = -\mu^2$. Defining the particle velocity by 
{\bf v} = {\bf p}/E the formulae for its energy and momentum become:
\begin{equation}
E = \frac{\mu}{\sqrt{v^2 -1}}
\end{equation} 
\begin{equation}
{\bf p} = \frac{\mu {\bf v}}{\sqrt{v^2 -1}}
\end{equation} 
Thus, the energy and 3-momentum of the faster-than-light particle are always 
real. As $v$ approaches 1 both the energy and momentum unlimitedly grow. 
Contrary, with the velocity increase they decrease, the energy 
approaching to zero at $v$ approaching to infinity, and the 3-momentum tending
to the finite value $\mu$. The sign of the energy can be changed by a suitable
Lorentz transformation,   
\begin{equation}
E~^\prime = \frac{E-{\bf pu}} {\sqrt{1 - u^2}} =\frac{E(1-{\bf vu})} 
{\sqrt{1 - u^2}},
\end{equation}
if ${\bf v u} > 1$, where ${\bf u}$ is the relative velocity of two reference 
frames. Simultaneously the sign of the time component of the particle world
line is changed. A coherent explanation of these changes was suggested in
\cite{bds}, denoted as {\em the principle of reinterpretation}. Accordingly to 
this principle, a faster-than-light particle of negative energy moving 
backward in time should be interpreted as an antiparticle of positive energy 
moving forward in time and in the opposite spatial direction. This 
reinterpretation is analogous to that proposed by Dirac, St\"uckelberg, 
Wheeler and Feynman for positrons as negative energy electrons going backward 
in time \cite{dirac,stuck,feyn}.
    
\section{Cherenkov radiation by tachyons}
\setcounter{equation}{0}
\renewcommand{\theequation}{3.\arabic{equation}}
If tachyons possess electric charge they should radiate Cherenkov
radiation since the laws of classical electrodynamics are also valid for
charged faster-than-light particles, see \cite{lienwi}.
The cone angle of the Cherenkov radiation by tachyons $\theta_c$ 
is related to the tachyon velocity $v$ and to the radiator refraction index $n$ 
in the same way as for ordinary particles:   
\begin{equation}
\cos \theta_c = \frac{1} {n v}.
\end{equation}
The validity of the formula (3.1) for tachyons follows from the fact that this
formula has purely kinematic origin. It can be obtained from the kinematics 
of the reaction
\begin{equation}
   t \rightarrow t^\prime + \gamma,
\end{equation}
where $t$ designates a charged tachyon, by use, for example, of 
the equation of four-momentum conservation:  
\begin{equation}
   P = P^\prime + K,
\end{equation}
where $P, P^\prime$ are tachyon four-momenta before and after emission of a
Cherenkov photon, respectively, and $K$ is a four-momentum of the photon.
Moving $K$ to the left side of the equation (3.3) and squaring both sides of it
we get
\begin{equation}
   (P - K)^2 = (P^\prime)^2,
\end{equation}
which reduces to
\begin{equation}
    (P K) = E \omega  -{\bf p k} = 0,    
\end{equation}
where $E$ and $\omega$ are energies of the initial tachyon and the Cherenkov
photon, respectively, and ${\bf p}$, $\bf{k}$ are their 3-momenta. For photons
of optical and near-optical frequencies, propagating in medium, the relation
between $\omega$ and $k$ is given by \cite{jackson}
\begin{equation}
    \omega ~n(\omega) = k(\omega),    
\end{equation}  
where $n(\omega)$ is the refraction index of the medium. Then (3.5) transforms
to
\begin{equation}
    E  - p \cos \theta_c ~n(\omega) = 0,    
\end{equation}
from which (3.1) follows, taking into account that tachyon velocity $v$
equals to $p/E$. 
 

In spite of the similar kinematics, there is a drastic distinction 
of the dynamics of the tachyon Cherenkov radiation 
from that of ordinary charged particles. The latter operates with a 
spectrum of the radiation frequencies, $\omega$, restricted by a 
narrow $\omega$ band in which the refraction index of the medium
passed by the particle, $n(\omega)$, is greater than $c/u$, where 
$u$ is the particle velocity.
This determines the Cherenkov radiation from the ordinary particles to be 
restricted within optical and near ultraviolet regions. For tachyons the 
Cherenkov radiation condition is satisfied at any radiation frequency even in 
the vacuum. As a result, a straightforward extrapolation of ordinary particle
Cherenkov radiation to the tachyonic case leads to an infinite Cherenkov energy 
loss \cite{cawley,sengupta}, and the only definite prediction which can be made
for the tachyon Cherenkov radiation in this case is the characteristic angle of
the radiation defined by a formula (3.1) since it has a purely kinematic origin.

Therefore first of all one has to formulate general principles in a frame of 
which the Cherenkov radiation of tachyons has to be considered. This 
consideration has to be carried out within a Lorentz-non-invariant approach 
to the tachyon Cherenkov radiation,
which is a mandatory condition for any tachyon theory \cite{pvcaus,ttheor}. 
This restricts the maximum frequency of the tachyon Cherenkov radiation in
the lab system to the so called quantum limit, $\omega_{max} = E_t$,
where $E_t$ is the tachyon energy in this system. Further, the problem of the 
intrinsic size of a tachyon has to be addressed. In order to understand why
this problem is very important in the tachyonic case, it is worthwhile to 
consider a hypothetical Cherenkov radiation from spinless (scalar) tachyons.

Scalar tachyons have to possess some kind of spherical symmetry. Let us 
consider first the case of a point-like tachyon. As was estimated in 
\cite{alv1}, the Cherenkov energy loss of point-like charged tachyons per 
unit length is extremely high converting the tachyons into virtual, rather 
than free, particles. 
This induces an idea to consider the tachyons possessing finite sizes.
Therefore the question about a tachyon form-factor, i.e. the question about the
tachyon size and the shape of its charge distribution, appears in any attempt 
of the realistic calculation of the tachyonic Cherenkov radiation. 
Interestingly, it has been understood a long time ago by A. Sommerfeld who 
considered, before special relativity  appeared, the radiation from an 
electron moving in vacuum with a superluminal speed \cite{sommer1,sommer2}.
His estimations of the energy loss by such an electron due to this radiation 
contain a characteristic size of the electron, $a_0$: 
\begin{equation}
\frac{dE}{dx} = -\frac{9 e^2 (v^2 - 1)}{4 v a_0^2},
\end{equation}
and are close to the 
results of modern calculations for the Cherenkov radiation by a finite-size,
``spherically symmetric" charged tachyon, see \cite{jones,rock,fain}.
In the case of a scalar tachyon its intrinsic size can be 
characterized by a single parameter only, say, $a_0$, with the visual 
appearance of the tachyon longitudinal size affected by a Lorentz contraction 
to be $a = a_0 \sqrt{v^2 - 1}$. 
Then the ``spherical shape" of such a tachyon would be achieved at a rather 
strange value of its velocity, $v = \sqrt{2}$, which has no particular 
meaning among all possible tachyon velocities, $1 < v < \infty$. Thus we see 
that the hypothesis of a scalar tachyon, considered from the point of view of
its form-factor, looks rather unnatural \footnote{Moreover, as noticed in 
\cite{ttheor}, this hypothesis fails on the observational ground since the
existence of scalar tachyons would lead to the instability of photons.}.

On the other hand, the situation becomes quite natural in the case of 
consideration of tachyons as being 
realizations of the UIR's of discrete series $D_s^+$ and $D_s^-$ of
the Poincar\'{e} group, as suggested in \cite{ttheor}. 
Such tachyons possess axially-symmetric form-factors 
characterized by two parameters, $\rho$ and $l_0$, 
the former being associated with the 
transversal tachyon size, and the latter with the longitudinal one. The 
preferred hierarchy of sizes seems to be $l_0 >> \lambda \geq \rho $, where 
$\lambda$ is the tachyon Compton length, $1/\mu$, and $\rho$ may be vanishingly 
small. With such a hierarchy the Cherenkov energy loss in the classical limit 
is determined by the parameter $l_0$ only \cite{lienwi,pvcher}, 
and can be expressed by a formula  
\begin{equation}
\frac{dE}{dx} = -f \frac{2 e^2}{l_0^2},
\end{equation}
where $f$ is a factor depending on the model of the tachyon form-factor.
Several models of the tachyon axially-symmetric form-factors, including one
with non-zero parameter $\rho$, were considered in \cite{pvcher}, and the
factor $f$ was found to be of order 1 varying by an order of magnitude.
A much bigger uncertainty comes from the indefiniteness of the parameter 
$l_0$, which may lie in the range of $10^{-12} - 10^{-10}$~cm \cite{ttheor}. 
However, even in the case of $l_0 \approx 10^{-10}$~cm the classical 
Cherenkov energy loss by a high energy charged tachyon would be very high, 
exceeding 2~GeV per $\mu m$.

Fortunately, the situation changes with a quantum-mechanical approach\footnote{
The situation is similar to the case of the classical radiation of an atomic
electron which would lead to a high energy loss by the electron through the
synchrotron radiation due to electron orbital motion. As well known
from quantum mechanics, such a radiation is absent 
in the case of the atomic ground state, and the radiation has a 
discrete spectrum in the case of exited electron orbits, as a 
consequence of quantum-mechanical selection rules.}. It turns out that 
the Cherenkov radiation by tachyons in the high energy $\gamma$ range is 
strongly affected by selection rules of angular momentum conservation.
In the vacuum such a radiation is strongly suppressed for scalar tachyons and 
for tachyons of the minimal helicity, $|h| = 1/2$. For higher helicity tachyons 
the energy of a radiated Cherenkov $\gamma$ is tightly restricted by the
relation, obtained within the quasi-classical approach:
\begin{equation}
E_\gamma = \frac{\sqrt{(4h^2 -1)~p^2 - \mu^2} -E}{2h^2 -1},
\end{equation} 
where $h$ is a tachyon helicity, and $p$ and $E$ are 
the tachyon 3-momentum and energy, respectively.
An emission of the next Cherenkov $\gamma$ is governed by the same relation.   
This leads to the discreetness of a single tachyon radiation spectrum 
and to the suppression of the tachyon Cherenkov radiation intensity 
by several orders of magnitude. An accurate estimation of this
suppression depends on the quantum-mechanical widths of the spectrum lines 
which, unfortunately, are not calculated yet. 
This prevents making definite predictions for the tachyon 
radiation intensity in the high energy $\gamma$ range.
   
On the other hand, the question about the tachyon behaviour in standard 
Cherenkov detectors is much more clear. At low radiation frequencies 
corresponding to optical and near ultraviolet regions, where these detectors 
are sensitive, the spectrum of the tachyon Cherenkov radiation 
is expected to be classical, i.e. continuous, 
and with quite loose assumptions about the tachyon intrinsic size 
$l_0$~\footnote{Just excluding macroscopic scale for $l_0$.} it is predicted
to have the shape of $\omega d\omega$, similarly to that of the radiation from
ordinary particles \cite{pvcher}. 
Then the spectrum of $detected$ Cherenkov 
photons will be determined by the condition of the optical transparency of the 
radiator, convoluted with the quantum efficiency of the detector of the 
radiation. For example, in the case of the DELPHI Barrel RICH the spectrum 
of detected photons extends from 5.6~eV to 7.5~eV \cite{rich}. 

The number of Cherenkov photons in this region is also expected, in the
classical limit, to be very close to that from ordinary relativistic particles.
However these expectations can fail quantum-mechanically, at angles essentially
exceeding the minimal Cherenkov angles (defined by $\cos\theta_c^{min} = 1/n$), 
i.e. at wide angles corresponding to $v >> 1$. The result could be 
a violation of the $\sin^2 \theta_c$ law for the Cherenkov radiation intensity
even at low radiation frequencies, especially in low density (gaseous) 
Cherenkov radiators. 

\section{Ionization loss of tachyons}
\setcounter{equation}{0}
\renewcommand{\theequation}{4.\arabic{equation}}
In the frame of our assumptions (standard electromagnetic interaction 
of tachyons) the Lorentz force acting on a charged tachyon 
induced by an atomic electron of an atom traversed by the tachyon 
(we neglect the atomic magnetic field) is reduced to 
\begin{equation}   
{\bf F} = e{\bf E},
\end{equation}  
where $e$ is the tachyon (presumably unit) charge 
and {\bf E} is the electric field of the electron which can be 
approximated by the Coulomb field since the electron velocities are much 
smaller than $c$. Thus, at this approximation, the force between the tachyon 
and the electron is:
\begin{equation}   
{\bf F} = \frac{e^2} {r^3} {\bf r}~.
\end{equation}   
The momentum transferred to the electron equals the time integral over 
the force acting in the direction perpendicular to that of the tachyon motion:
\begin{equation}   
\Delta  p_e = \int_{-\infty} ^{+\infty} F_\bot dt = 
e^2 \int_{-\infty} ^{+\infty} \frac{b dt}{[b^2+(vt)^2]^{3/2}} = \frac{2e^2}{vb}.
\end{equation}
Here $v$ is a tachyon velocity and $b$ is a tachyon impact parameter. 
The energy acquired by the electron equals $\Delta  p_e^2 /2 m_e$. 
Thus the tachyon ionization loss is expected to be proportional to $1/v^2$, 
similarly to that of ordinary particles (which is natural since 
the ionization mechanisms are identical in both cases).
Furthermore, it has to grow logarithmically with $v \rightarrow 1$ due to 
Lorentz contraction of the tachyon electric field which leads to an
enhancement of the tachyon field strength at the periphery of the impact 
parameter space, $dE/dx \sim \ln (\gamma^2)/v^2$.  
More information about this growth can be
found in textbooks on the classical electrodynamics. Though the ionization
loss by a charged tachyon is not considered in these textbooks, their formulae
derived for ultrarelativistic particles (having velocities close to $c$)
are applicable to the tachyon case also. Thus, for a tachyon with a
velocity $v \approx 1$
\begin{equation}
\frac{dE}{dx}\Big{(}\gamma\Big{)} = \frac{dE}{dx}\Bigg{|}_0~
\Bigg{(} \ln \frac{2m_e \gamma^2}{I} - 1 \Bigg{)}, 
\end{equation}
where $\gamma = 1/\sqrt{v^2-1}$ is a tachyon Lorentz factor,
$\frac{dE}{dx}\Big{|}_0$ is the minimum ionization loss of a particle in a
given medium (the mip), $m_e$ is the electron mass, and $I$ is a mean atomic
excitation energy. Also, the density effect and the Fermi plateau are
expected as usual.

For low energy tachyons (i.e for tachyons with $v >> 1$) the ionization loss 
is expected to drop faster than $1/v^2$, as has been noted in~\cite{pvcher}.  
  
\section{Tachyon energy loss in electromagnetic calorimeters}
\setcounter{equation}{0}
\renewcommand{\theequation}{5.\arabic{equation}}
Electromagnetic calorimeters are designed to measure the energy and
position of electromagnetic showers produced by high energy electrons 
(positrons) and photons. These showers have certain characteristics which 
allow them to be distinguished from the electromagnetic calorimeter response
to other charged particles, such as muons, pions, kaons and high energy protons.
The latter leave in these calorimeters, in most of the cases, only a part
of their energy due to ionization. The creation of electromagnetic showers
by electrons and positrons (and by photons after their conversion to 
electron-positron pairs) is determined by their strong energy loss in
the electromagnetic calorimeters due to bremsstrahlung radiation.

Tachyon behaviour in the electromagnetic calorimeters is expected to be very
similar to that of high energy electrons, and not to other charged particles.
In order to understand this let us consider first the formula for bremsstrahlung
radiated by electrons and positrons colliding at high energy (we have taken
electron-positron collisions in order to avoid the identity of interacting 
particles). 

The production rates for the bremsstrahlung photons from
colliding $e^+ e^-$ (initial state radiation) and from final $e^+ e^-$ 
(final state radiation) in the soft photon region can be calculated at once 
using an universal formula (see e.g. \cite{muonbrems}, where a similar formula
has been applied to the calculation of the bremsstrahlung rate in the reaction
$e^+ e^- \rightarrow \mu^+ \mu^-$):
\begin{equation}
\frac{dN_{\gamma}}{d^{3}\vec{k}}
=
\frac{\alpha}{(2 \pi)^2} \frac{1}{E_\gamma}
\int d^3 \vec{p}_{e^+} d^3 \vec{p}_{e^-}  
\sum_{i,j} \eta_{i} \eta_{j}
\frac{(\vec{p}_{i \bot} \cdot \vec{p}_{j \bot}) }{ ( P_{i} K )  ( P_{j} K )}
\frac{ d N_{e^+}}{ d^{3} \vec{p}_{e^+}  } 
\frac{ d N_{e^-}}{ d^{3} \vec{p}_{e^-}  } 
\end{equation}

\noindent
where $K$ and $\vec{k}$ denote photon four- and three-momenta,
$P$ are the 4-momenta of $e^+, e^-$, and $\vec{p}_e$ are their 3-momenta,
while their transversal (w.r.t. the photon direction) momenta are denoted by
$\vec{p}_{i \bot} = \vec{p}_i-(\vec{n} \cdot \vec{p}_i) \cdot \vec{n}$, where
$\vec{n}$ is the photon unit vector, $\vec{n} = \vec{k}/k $;
$\eta=1$ for the initial $e^-$ and for the outgoing $e^+$,
$\eta=-1$ for the initial $e^+$ and for the outgoing $e^-$,
and the sum is extended over all (initial and final) electrons and positrons;
the last two factors in the integrand are the final electron and positron
differential spectra.

In 3-vector form, the denominator of formula (5.1) contains terms of type of 
$(1 - v \cos \theta_\gamma)$, 
where $v$ is the particle velocity and $\theta_\gamma$
is the photon emission angle. For relativistic electrons  $v \approx 1$ and
typical values of $\theta_\gamma$ are of order of $1/\Gamma$, $\Gamma$ being 
the electron Lorentz-factor, quite big for relativistic electrons. 
For example, electrons at LEP1 have $\Gamma \approx 10^{5}$, and for them
$(1 - v \cos \theta_\gamma) \approx 10^{-10}$. The extreme smallness of these
denominator terms is called {\em collinear singularity}. It determines the
high bremsstrahlung rate from high energy $e^+, e^-$ and, in turn,
their showering in the electromagnetic calorimeters. 

For tachyons, due to their $v > 1$ there always exists an angle $\theta_\gamma$
(even not very small) for which $1 - v \cos\theta_\gamma \approx 0$, thus 
satisfying the collinear singularity condition, i.e. ensuring a high 
bremsstrahlung rate, which is practically independent of the tachyon mass 
(unless the tachyon becomes non-relativistic, see remark below). Therefore 
one can expect that the energy loss of relativistic tachyons in electromagnetic 
calorimeters is very similar to that of the high energy electrons. 
This may not be true for non-relativistic tachyons (having $v >> 1$) 
since the collinear singularity condition requires wide angles of the radiation
at $v >> 1$, which is expected to be suppressed by the quantum effect of  
angular momentum conservation, analogously to the similar effect in the case 
of the wide angle Cherenkov radiation, mentioned in Sect.~3. 

\section{Tachyons in transition radiation detectors}
Transition radiation detectors (TRD's) are used in high energy physics 
experiments as particle identificators (mainly for the separation of electrons 
and pions), and as tracking and trigger devices. The particle identification 
properties of TRD's are realized when highly relativistic charged particles 
with the Lorentz factors $\gamma \geq 10^3$ cross many interfaces of two media
with different refractive indices. However, at high particle momenta 
(corresponding to pion $\gamma$'s $\approx 100$) the 
pion/electron identification starts to deteriorate due to the relativistic 
rise of the specific loss of pions; thus the momentum range of the TRD's 
identification facility is spanned usually from 1 to 10~GeV/$c$, though in 
some cases it can be extended to higher momenta by an order of magnitude.   

Charged tachyons with the tachyon Lorentz factors 
$\gamma=1/\sqrt{v^2 -1} < 10^3$
are not expected to produce a significant amount of the transition radiation 
in TRD's, thus their energy loss in these detectors are expected to be 
dominated by ionization. Comparing the response of a TRD (similar to 
hadronic one in the case of tachyons) with the response of an electromagnetic 
calorimeter (similar, for tachyons, to the electron response, see Sect.~5) 
for a given particle one can use this comparison as an additional tachyonic 
signature when looking for tachyons with the mass parameters 
$\mu > 100$~MeV/$c^2$ in the high energy experimental data.  

\section{Conclusion}
Several experimental aspects of the tachyon hypothesis are considered in this 
note related to the expected behaviour of charged tachyons in a detecting
apparatus. In particular, the Cherenkov radiation by tachyons, their ionization
loss in tracking devices and that in electromagnetic calorimeters and TRD's
are considered. In summary, charged tachyons, if they exist, can be expected 
to behave in particle detectors (excepting TRD's) like high energy electrons, 
differing from the latter by anomalous ring Cherenkov radiation and, 
at high velocities, by anomalously low ionization.

\subsection*{Acknowledgements}
\vskip 3 mm
The author thanks Profs. K.~G.~Boreskov, F.~S.~Dzheparov, 
A.~A.~Grigoryan, O.~V.~Kancheli 
and Drs. O.~N.~Ermakov, B.~R.~French for fruitful discussions.


\newpage

\newpage
\begin{thebibliography}{ References}
\bibitem{wigner1} E.P. Wigner, Ann. Math. {\bf 40}, 149 (1939)
\bibitem{wigner2} E.P. Wigner, {\em Invariant Quantum Mechanical Equations of
Motion}, in a book {\em Theoretical Physics} (I.A.E.A. Vienna, 1963), p. 59 
\bibitem{bds} O.M.P. Bilaniuk, V.K. Deshpande, E.C.G. Sudarshan, Amer. J. Phys.
{\bf 30}, 718 (1962)
\bibitem{fein} G. Feinberg, Phys.Rev. {\bf 159}, 1089 (1967) 
\bibitem{newton} R.G. Newton, Phys. Rev., {\bf 162}, 1274 (1967)
\bibitem{roln} W.B. Rolnick, Phys. Rev. {\bf 183}, 1105 (1969)
\bibitem{parment} J.A. Parmentola, D.D.H. Lee, Phys. Rev. D {\bf 4}, 1912 (1971)
\bibitem{ein} A. Einstein, Ann. d. Phys. (Vierte Folge) {\bf 23}, 371 (1907). 
In this paper A.~Einstein noted that his theory (special
relativity at that moment) is not compatible with the possibility of 
faster-than-light signals.
\bibitem{tolman} R.C. Tolman, {\em The Theory of Relativity of Motion} (Univ.
of California Press, Berkeley 1917), p. 54 
\bibitem{moller} C. M{\o}ller, {\em The Theory of Relativity} (Oxford Univ. 
Press, 1952), p. 52
\bibitem{bohm} D. Bohm, {\em The Special Relativity} (Addison-Wesley Publishing
Company, New York, 1965), p. 155 
\bibitem{nielsen} H.B. Nielsen, {\em Tachyons in Field Theory}, in a book
{\em Tachyons, Monopoles and Related Topics}, E. Recami ed. (North-Holland 
Publishing Company, Amsterdam, 1978), p. 169   
\bibitem{pvcaus} V.F. Perepelitsa, {\em Causality, Relativity and 
Faster-Than-Light Signals}, in a book {\em Philosophical Problems of
Hypothesis of Superluminal Velocities} (Nauka Press, Moscow, 1986), p. 40
(in Russian)
\bibitem{ttheor} V.F. Perepelitsa {\em Looking for a Theory of 
Faster-Than-Light Particles}, arXiv:physics.gen-ph/14073245                  
\bibitem{dirac} P.A.M. Dirac, Proc. Cambr. Phil. Soc. {\bf 26}, 376 (1930)
\bibitem{stuck} E.C.G. St\"uckelberg, Phys. Rev. {\bf 74}, 218 (1948)
\bibitem{feyn} R.P. Feynman, Phys. Rev. {\bf 76}, 769 (1949)
\bibitem{lienwi} V.F. Perepelitsa, {\em Lienard-Wiechert potentials for
charged tachyons and several remarks on the tachyon Cherenkov radiation},
arXiv:hep-th/1502.0655 (2015)
\bibitem{jackson} J.D. Jackson, {\em Classical Electrodynamics} (John Willey
and Sons, Inc., New York, Chichester, Weinheim, Brisbane, Singapore, Toronto),
3rd edition, p.~325 (1999)
\bibitem{cawley} R.G. Cawley, Phys. Rev. D {\bf 2}, 276 (1970)
\bibitem{sengupta} N.D. Sen Gupta, Nucl. Phys. B {\bf 27}, 104 (1971) 
\bibitem{alv1} T. Alv\"ager, M.N. Kreisler, Phys. Rev. {\bf 171}, 1357 (1968) 
\bibitem{sommer1} A. Sommerfeld, Nachrichten K\"onigl. Gesellshaft 
Wissenshaften zu G\"ottingen, Math-Phys. Klasse, 1904, p. 406; 
~ibid, 1905, p.201
\bibitem{sommer2} A. Sommerfeld, Proc. Royal Acad. Amsterdam {\bf 7}, 346 
(1904); reproduced in A. Sommerfeld, {\em Gesammelte Schriften}, vol.~2 
(Braunschweig, 1968)
\bibitem{jones} F.C. Jones, Phys. Rev. D {\bf 6}, 2727 (1972) 
\bibitem{rock} E.B. Rockower {\em Generalized Cherenkov Radiation from 
Tachyons.} Dissertation. Waltham, Massachusetts (1975)
\bibitem{fain} M.I. Feingold, Theor. Math. Phys. {\bf 47}, 395 (1981) 
(in Russian)
\bibitem{pvcher} V.F. Perepelitsa, {\em Cherenkov Radiation of Extended 
Tachyon}. I. {\em Radiation characteristics}, ITEP 87-31 (1987); 
II. {\em Suggested experiments for tachyon search}, ITEP 87-34 (1987) 
(In Russian)
\bibitem{rich} W. Adam et al., Nucl. Instr. and Meth. A {\bf 367}, 233 (1995) 
\bibitem{muonbrems} DELPHI Collaboration, J. Abdallah et al.,
Eur. Phys. J. C {\bf 57}, 499 (2008)
\end{thebibliography}
\end{document}